\documentclass[useAMS,usenatbib]{mnras}
\usepackage{natbib}
\usepackage{graphicx}
\usepackage{epsfig}
\usepackage{amssymb}
\usepackage{amsmath}
\usepackage{natbib}
\usepackage{times}
\usepackage{subfigure}
\usepackage[hyphenbreaks]{breakurl}

\usepackage[english]{babel}  

\title[Stochastic perturbations in Keplerian flow]{Perturbations dynamics in Keplerian flow under external stochastic forcing}
\author[D. N. Razdoburdin]{D. N. Razdoburdin$^{1}$\thanks{E-mail:
d.razdoburdin@gmail.com} \\
$^{1}$Sternberg Astronomical Institute, Moscow M.V. Lomonosov State University, Universitetskij pr., 13, Moscow 119992, Russia}
\begin{document}

\date{
}

\pagerange{\pageref{firstpage}--\pageref{lastpage}} \pubyear{2020}

\maketitle

\label{firstpage}

\begin{abstract}
We investigate the dynamics of linear perturbations in Keplerian flow under external stochastic force.
To abstract from the details of flow structure and boundary conditions, we consider the problem in the shearing box approximation.
An external force is assumed to have zero mean, even so, induced perturbations form a steady-state, which provides angular momentum transfer to the periphery of the flow.
The most effective scenario is based on the transient amplification of induced vortices with the following emission of shearing sound wave, wherein the maximum of the flux linearly depends on Reynolds number.
Thus such a mechanism is significant for astrophysical flows, for which enormous Reynolds numbers are typical.
At the same time, addressing the problem analytically, we found that for incompressible fluid in the shearing box approximation stochastic forcing does not lead to average angular momentum transfer.
Thus the compressibility of the fluid plays an important role here, and one cannot neglect it.
\end{abstract}


\begin{section}{Introduction}
Starting from the pioneering papers \cite{shakura-1972} and \cite{shakura-sunyaev-1973}, the mechanism giving rise to effective viscosity and the related angular momentum transfer in the accreting flows is on the agenda. The application of magnetorotational instability (\cite{velikhov-1959, chandrasekhar-1960}) to accretion theory by \cite{balbus-hawley-1991, hawley-balbus-1992, balbus-hawley-1992} provided a powerful mechanism of effective viscosity origin due to the turbulization of the flow. However MRI requires high gas ionization as well as the initial magnetic field (see \cite[section 1]{mukhopadhyay-2005} for a more detailed discussion). Another branch of turbulization mechanisms, such as vertical shear instability \cite{nelson-2013}, subcritical baroclinic instability \cite{lesur-papaloizou-2010} and zombie vortex instability \cite{marcus-2015}, provide dimensionless angular momentum flux of only about $\alpha \sim 10^{-3}$ (for a more detailed review with respect to protoplanetary disks see \cite[section 1.5]{armitage-kley-2019}).

So, in the current paper we look back onto a purely hydrodynamic model but take into account that accretion disks are not isolated from their surrounding.
We approximate the action of the surrounding as an external stochastic force with zero mean value (a discussion of possible sources of forcing can be found in \cite [section 4]{ioannou-kakouris-2001}).
In such a model, external action induces perturbations capable of transient amplification.
The main aim of the current paper is to study the possibility of angular momentum transfer by them.

A similar approach was previously successfully applied while investigating the rotating Couette flow by \cite{chen-1987}, channel flow  by \cite{bamieh-dahleh-2001} and plane Couette flow by \cite{farrell-ioannou-1993} and \cite{khujadze-2006}.
In the astrophysical context, \cite{farrell-ioannou-1999} used similar method for investigating large scale magnetic fields generation.
Angular momentum flux that occurs due to the generation of perturbations by external stochastic forcing was investigated in \cite{ioannou-kakouris-2001} for the 2D incompressible fluid.

In the current paper we set the goal to generalize the results of \cite{ioannou-kakouris-2001} by taking compressibility and vertical spatial dimension into account.
At the same time, we reduce the model of flow structure to the shearing box.
Such approximation allows us to avoid the influence of boundary conditions and global flow structure, which are model dependent.
Another advantage of the shearing box model consists in the possibility of transit from coordinate description to spatial Fourier harmonics (SFH) in dynamical equations. 
This allows not only to simplify the equations significantly (in SFH description we get ordinary differential equations instead of equations in partial derivatives) but also to analyze the resulting solutions in more detail.
On the other hand, for a single SFH non-trivial steady state solution with external action does not exist.
For that reason, the additional procedure of transition from non-steady state solution for single SFH to steady state for the ensemble of harmonics is described.

This paper is organised as follows.
In section \ref{sec::equations} we present the equation for the dynamics of single SFH under stochastic forcing.
The way of transition from the dynamics of the single SFH to the spectrum of perturbations in the flow is described in section \ref{sec::spectra}.
An analytical solution for the subcase of small scale 2D perturbations is introduced in section \ref{sec::subcase}.
In section \ref{sec::results} we collect the results of both numerical and analytical calculations and analyze them, and in section \ref{sec::summary} we sum up the results of the investigation.

\end{section}

\begin{section}{Equations for single spatial fourier harmonic}{\label{sec::equations}}

We solve the equation for Eulerian perturbations in shearing box approximation, i.e. in local Cartesian frame corotating with angular velocity $\Omega_0$ (see \cite{goldreich-lynden-bell-1965}, \cite{umurhan-regev-2004}).
We neglect details of energy balance in the fluid and adopt polytropic equation of state; this allows us to use continuity equation for enthalpy perturbation $W$.
Dissipation is described both by kinematic and bulk viscosity coefficients.
External stochastic forcing is represented by heterogeneous terms $\zeta_{x, y, z, w}$ in the dynamic equations.

\begin{equation}{\label{eq::ux}}
\left (\frac{\partial}{\partial t} - q\Omega_0 x\frac{\partial}{\partial y} \right ) u_x - 2\Omega_0 u_y
+ \frac{\partial W}{\partial x} - h_x - g_x = \zeta_x,
\end{equation}

\begin{equation}{\label{eq::uy}}
\left ( \frac{\partial}{\partial t} - q\Omega_0 x\frac{\partial}{\partial y} \right ) u_y + 
(2 - q)\Omega_0 u_x 
+\frac{\partial W}{\partial y} - h_y - g_y = \zeta_y,
\end{equation}

\begin{equation}{\label{eq::uz}}
\left ( \frac{\partial}{\partial t} - q\Omega_0 x\frac{\partial}{\partial y} \right ) u_z
+\frac{\partial W}{\partial z} - h_z - g_z = \zeta_z,
\end{equation}

\begin{equation}{\label{eq::w}}
\left ( \frac{\partial}{\partial t} - q\Omega_0 x\frac{\partial}{\partial y} \right ) W + 
c_s^2 \left (\frac{\partial u_x}{\partial x} + \frac{\partial u_y}{\partial y} + \frac{\partial u_z}{\partial z} \right) = \zeta_w,
\end{equation}

\begin{equation}{\label{eq::visc}}
h_{x, y, z} = \nu \left ( \frac{\partial^2}{\partial x^2} + \frac{\partial^2}{\partial y^2} + \frac{\partial^2}{\partial z^2}\right ) u_{x,y,z}, 
\end{equation}

\begin{equation}{\label{eq::bulk}}
g_{x, y, z} = (\nu_b + \nu/3) \frac{\partial}{\partial x,\partial y, \partial z} \left ( \frac{\partial u_x}{\partial x} + \frac{\partial u_y}{\partial y} + \frac{\partial u_z}{\partial z} \right ),
\end{equation}

\begin{equation}{\label{eq::zeta:w}}
\zeta_w = 0,
\end{equation}
where $u_x$, $u_y$, $u_z$ and $W$ are Eulerian perturbations of velocity components and enthalpy, $\zeta_{x, y, z}$ are components of external stochastic force.
Component $\zeta_w$ represents stochastic addition to continuity equation.
In the present paper we always set $\zeta_w = 0$ (we add such a variable to the equation, in spite of its equality to zero to further rewrite the set in matrix form).
Shear rate $q$, rotation frequency $\Omega_0$, sound speed $c_s = H / \Omega_0$, half thickness of the flow $H$, kinematic viscosity $\nu$ and bulk viscosity $\nu_b$ are constants.

In the current study, we focus on the case of Keplerian rotation, so the shear rate is always equal to Keplerian $q = 3/2$.
Below we always use dimensionless form of equations (\ref{eq::ux} -- \ref{eq::bulk}).
This means that we transit to new coordinates
\begin{equation}
x^{\prime} = x/H, y^{\prime} = y/H, z^{\prime} = z/H, t^{\prime} = \Omega_0 t.
\end{equation}
Moreover, we made parameterization of viscosity coefficients by dimensionless Reynolds numbers:
\begin{equation}
R \equiv H^2\Omega_0 / \nu 
\end{equation}
\begin{equation}
R_b \equiv H^2\Omega_0 / \nu_b
\end{equation}

Following \cite{ioannou-kakouris-2001}  we represent each component of $\mathbf{\zeta}$  as a linear combination of spatially $\delta$-correlated, temporally Gaussian stochastic process $\mathbf{\eta}$.
Each of the processes $\mathbf{\eta}$ has zero ensembles mean.
\begin{equation}
\zeta_{i} = \sum \limits_{k} f_{ik} \eta_k,
\end{equation}
\begin{equation}{\label{forcing_covariance}}
\begin{aligned}
&\left<\eta_i(x_1, y_1, z_1, t_1), \eta_j^*(x_2, y_2, z_2, t_2)\right> = \\
& = \delta_{ij} \delta(x_1 - x_2) \delta(y_1 - y_2) \delta(z_1 - z_2) \delta(t_1 - t_2),
\end{aligned}
\end{equation}
\begin{equation}
\left<\eta_j(x, y, z, t)\right> = 0
\end{equation}
(here running indexes $i, j, k$ denote one of the components ${x, y, z, w}$).
Such representation of external force allows to set non-zero covariance between force's components
\begin{equation}{\label{eq::covariance}}
\begin{aligned}
&\left<\zeta_i(x_1, y_1, z_1, t_1), \zeta_j^*(x_2, y_2, z_2, t_2)\right> = \\
& = \left(\sum\limits_{k=x}^{z} f_{ik}f_{jk}\right) \delta(x_1 - x_2) \delta(y_1 - y_2) \delta(z_1 - z_2) \delta(t_1 - t_2),
\end{aligned}
\end{equation}
and at the same time, guarantees their zero ensemble mean
\begin{equation}
\left<\zeta_j(x, y, z, t)\right> = 0.
\end{equation}

The covariance properties of the external forcing have a great influence on the induced perturbations. 
We denote the matrix of forcing covariance as follows:
\begin{equation}{\label{eq::FFdag:intro}}
(\mathbf{FF}^{\dag})_{ij} = \sum\limits_{k} f_{ik}f_{jk}.
\end{equation}
Note that since $\zeta_w = 0$ the corresponding column and row of $\mathbf{FF}^{\dag}$ are vanished.

A common way (starting from \cite{goldreich-lynden-bell-1965}) for solving the equations for perturbations in shearing box consists in transition to spatial Fourier harmonic (SFH).
We made such a transition by integral transformation:
\begin{equation}{\label{eq::fourie}}
\begin{pmatrix}
u_x    \\
u_y    \\
u_z    \\
W      \\
\eta_x \\
\eta_y \\
\eta_z \\
\eta_w
\end{pmatrix} = \int \begin{pmatrix}
\hat{u}_x(t)    \\
\hat{u}_y(t)    \\
\hat{u}_z(t)    \\
\hat{W}(t)      \\
\hat{\eta}_x(t) \\
\hat{\eta}_y(t) \\
\hat{\eta}_z(t) \\
\hat{\eta}_w(t)
\end{pmatrix} \exp\left({\rm i} \left[\tilde{k}_x x + k_y y + k_z z\right]\right) d\mathbf{k}^3,
\end{equation}

Here $\tilde k_x$ is the x-component of the wavevector, which is changed in time due to the shear
\begin{equation}{\label{eq::kx}}
\tilde k_x = k_x + q k_y t.
\end{equation}

Fourier amplitudes of external force
\begin{equation}
\hat{\zeta}_i = \sum\limits_k f_{ik}\hat{\eta_k}
\end{equation}
have zero ensembles mean

\begin{equation}
\left< \hat{\zeta}_{x, y, z, w}(t)\right> = 0,
\end{equation}
and their covariance equals to 
\begin{equation}
\left< \hat{\zeta}_i(t_1)\hat{\zeta}_j^*(t_2)\right> = (\mathbf{FF}^{\dag})_{ij} \delta(t_1 - t_2).
\end{equation}

Now we can write equations for spatial Fourier harmonics under external stochastic forcing:
\begin{multline}{\label{eq::SFH_x}}
\frac{d \hat u_x}{d t} = 2\hat u_y - {\rm i}\tilde k_x \hat W - R^{-1} k^2 \hat u_x - \\
    (R^{-1}/3 + R_b^{-1}) \tilde k_x (\tilde k_x \hat u_x + k_y \hat u_y + k_z \hat u_z) + \\
    f_{x, x} \mathbf{\hat\eta}_x + f_{x, y} \mathbf{\hat\eta}_y + f_{x, z} \mathbf{\hat\eta}_z + f_{x, w} \mathbf{\hat\eta}_w,
\end{multline}

\begin{multline}{\label{eq::SFH_y}}
\frac{d \hat u_y}{d t} = -(2-q)\hat u_x - {\rm i} k_y \hat W - R^{-1} k^2 \hat u_y - \\
    (R^{-1}/3 + R_b^{-1}) k_y (\tilde k_x \hat u_x + k_y \hat u_y + k_z \hat u_z) + \\
    f_{y, x} \mathbf{\hat\eta}_x + f_{y, y} \mathbf{\hat\eta}_y + f_{y, z} \mathbf{\hat\eta}_z + f_{y, w} \mathbf{\hat\eta}_w,
\end{multline}

\begin{multline}{\label{eq::SFH_z}}
\frac{d \hat u_z}{d t} = - {\rm i} k_z \hat W - R^{-1} k^2 \hat u_z - \\
    (R^{-1}/3 + R_b^{-1}) k_z (\tilde k_x \hat u_x + k_y \hat u_y + k_z \hat u_z) + \\
    f_{z, x} \mathbf{\hat\eta}_x + f_{z, y} \mathbf{\hat\eta}_y + f_{z, z} \mathbf{\hat\eta}_z + f_{z, w} \mathbf{\hat\eta}_w,
\end{multline}

\begin{multline}{\label{eq::SFH_W}}
\frac{d \hat W}{d t} = - {\rm i}\, ( \, \tilde k_x \hat u_x + k_y \hat u_y + k_z \hat u_z\,) + \\
    f_{w, x} \mathbf{\hat\eta}_x + f_{w, y} \mathbf{\hat\eta}_y + f_{w, z} \mathbf{\hat\eta}_z + + f_{w, w} \mathbf{\hat\eta}_w,
\end{multline}
where $k^2 = \tilde{k}_x^2 + k_y^2 + k_z^2$.

Note again that we save $w$-component of forcing only for rewriting the equation set in matrix form, thus the corresponding components of $f_{i, j}$ are always vanished: $f_{i, w} = f_{w, i} = 0$.
The matrix form of that set of equations is quite simple:
\begin{equation}{\label{eq::dynamic}}
\frac{d \hat{\mathbf{q}}}{dt} = \mathbf{A \hat{q}} + \mathbf{F \hat \eta}, 
\end{equation}
where $\hat{\mathbf{q}} = \{\hat u_x, \hat u_y, \hat u_z, \hat W\}$ is state vector, and $\mathbf{A}$ is matrix of the dynamical operator for noiseless case.
For perturbations in SFH-representation operator $\mathbf{A}$ is time dependent, i.e. it is non-autonomous (see \cite{farrell-ioannou-1996b} for a more detailed description of non-autonomus operators).

One can decompose the set (\ref{eq::SFH_x} -- \ref{eq::SFH_W}) for complex-valued state vector $\mathbf{\hat q}$ into two independent equivalents: real-valued equation for state vectors $\mathbf{\hat q}^R = \{\Re \hat u_x, \Re \hat u_y, \Re \hat u_z, \Im \hat W\}$ and $\mathbf{\hat q}^I = \{\Im \hat u_x, \Im \hat u_y, \Im \hat u_z, -\Re \hat W\}$.
Thus we can solve it only for one of the real-valued state vectors (that property significantly simplifies numerical solver).

Following \cite{ioannou-kakouris-2001} and \cite{farrell-ioannou-1999} we solve the equation (\ref{eq::dynamic}) for covariance matrix $\mathbf{C}$
\begin{equation}{\label{eq::Cdef}}
\mathbf{C}_{ij}(t) = \left<\mathbf{\hat{q}}_i(t), \mathbf{\hat{q}}_j(t)\right>,
\end{equation}
where brackets denote averaging over an ensemble of realizations.

The dynamic equation for matrix $\mathbf{C}$ is the same as was found in the previous studies (\cite{farrell-ioannou-1996a}, \cite{farrell-ioannou-1999}, \cite{ioannou-kakouris-2001}) for the case of authonomus operators:
\begin{equation}{\label{eq::Cdyn}}
\frac{d\mathbf{C}}{dt} = \mathbf{FF}^{\dag} + \mathbf{AC} + \mathbf{CA}^{\dag}
\end{equation}
That is the so-called differential Lyapunov equation.
We place its detailed derivation for the case of non-autonomous operators in appendix \ref{appendix::C::derivation}.
The explicit view of $\mathbf{A}$ and $\mathbf{A}^{\dag}$ can be found in appendix \ref{appendix::Adag::derivation}.

Thereby in the course of the transition from equation (\ref{eq::dynamic}) to equation (\ref{eq::Cdyn}), we lost information about the values of perturbation's Fourier amplitudes: from equation (\ref{eq::Cdyn}), we can derive only the values of ensemble-averaged quadratic forms.
On the other hand, we got rid of stochastic terms in the equation.
Such an exchange looks quite reasonable since both the quantities that interest us (energy and angular momentum flux) can be derived from matrix $\mathbf{C}$:
\begin{equation}{\label{eq::e}}
e(t) = \left< \hat u_x^2 + \hat u_y^2 + \hat u_z^2 + \hat W^2 \right> = \mathrm{trace}~\mathbf{C},
\end{equation}
\begin{equation}
\phi(t) = \left< \hat u_x \hat u_y \right>  = \mathbf{C}_{x, y}.
\end{equation}

Combining the equations (\ref{eq::Cdyn}) and (\ref{eq::e}) we get the following expression for $\dot e$:
\begin{equation}{\label{eq::dot_e}}
\dot e = \mathrm{trace}~\mathbf{FF}^{\dag} + \mathrm{trace}~\left(\mathbf{AC} + \mathbf{CA}^{\dag}\right)
\end{equation}
The first term in the expression (\ref{eq::dot_e}) is associated with the energy injected in the flow by forcing, and the second one, in its turn, is associated with energy exchange processes due to the dynamical evolution of perturbation.
We denote the injected energy as $\dot e_{in}$:
\begin{equation}{\label{eq::e_in}}
\dot e_{in} = \mathrm{trace}~\mathbf{FF}^{\dag}
\end{equation}

The equation (\ref{eq::Cdyn}) represents an ordinary differential equation for symmetric $4\times 4$ matrix (thus it is equivalent to the set of $10$ independent ODE).
To solve it we wrote a solver on C++ with the help of boost library \cite{boostLibrary}
\footnote{the source code is available by link \url{http://xray.sai.msu.ru/~dima/DynamicsOfKeperianFlowUnderStochasticForcing.html}}.
See also appendix \ref{appendix::solver::test} for the details of solver tests.

We pay most of the attention to the case of uncorrelated external forcing components.
That corresponds to the diagonal form of matrix $\mathbf{FF}^{\dag}$ (see expression \ref{eq::covariance}).
We also assume that the external force is isotropic, i.e $\mathbf{FF}^{\dag}_{x, x} = \mathbf{FF}^{\dag}_{y, y} = \mathbf{FF}^{\dag}_{z, z}$.
And since injected power represents natural normalization of the equations, we set $\dot e_{in} = 1$.
These three assumptions uniquely determine the form of the covariance matrix of forcing:
\begin{equation}{\label{eq::FUncorr3D}}
\mathbf{FF}^{\dag}_{uncorr} = 
\begin{pmatrix}
& 1/3 & 0   & 0   & 0 \\
& 0   & 1/3 & 0   & 0 \\
& 0   & 0   & 1/3 & 0 \\
& 0   & 0   & 0   & 0 \\
\end{pmatrix}.
\end{equation}

In the current paper we are interested in the steady state of perturbations under external forcing.
Such a state occurs when the action of external force is balanced by dissipations in the flow.
However, for single SFH steady state can only be a trivial:
\begin{equation}
\mathbf{C}^{\infty} = \lim \limits_{t\to \infty} \mathbf{C}(t) = 0,
\end{equation}
since viscous force becomes infinite for $\mathbf{k}^2 \to \infty$ (see expression \ref{eq::kx}).
Steady-state arises only when the whole spectrum of harmonics is considered. We describe the transition from non-steady solution for one SFH to a steady spectrum of perturbations in the next section.

In subsequent sections we will always deal with ensemble averaged quantities, thus we will no longer mention this.
\end{section}

\begin{section}{Steady state spectra of induced perturbation}{\label{sec::spectra}}
\begin{figure}
\includegraphics[width=\linewidth]{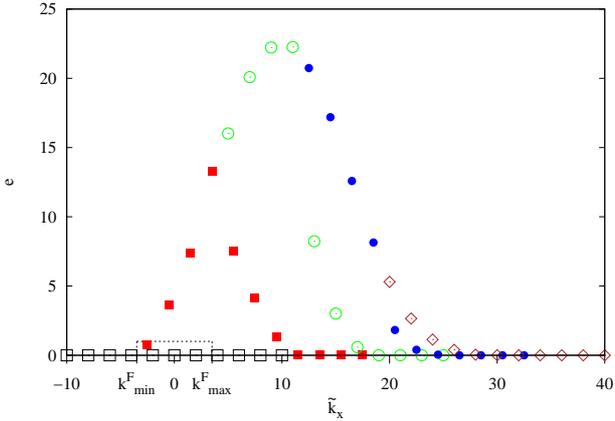}
\caption{
Energy and radial wavenumber evolution of the SFHs set is plotted here. Squares, filled squares, circles, filled circles and diamonds correspond to time moments $t = 0, 5, 10, 15, 20$. 
Initially, harmonics differ only by the value of $\tilde k_x$: from $\tilde k_x = -10$ to $\tilde k_x = 10$ with step $dk_x = 2$ for $t = 0$, initial values of covariance matrices are zero $C(0) = 0$ for all SFHs (thus, initial energy for all SFHs is zero).
The dashed rectangle denotes the spectral interval of the external forcing.
$k^F_{min} = -3.5$, $k^F_{max} = 3.5$, $k_y = 1, k_z = 0, R = 1000, R_b = \infty$.
}
\label{fig:NonSteadyStatePoints}
\end{figure}

In the previous section, we presented equations for the dynamics of single SFH.
The goal of the current one is to describe an algorithm for calculating steady-state spectra under temporary invariant forcing.

First of all, let us calculate the evolution of the set of SFH with zero initial condition.
This case corresponds to the turning-on of the forcing in an initially unperturbed flow.
Solving the equation (\ref{eq::Cdyn}) for each of the SFHs we calculate the evolution of covariance matrices.

As an example, we set uncorrelated forcing covariance matrix with unit injected power (equation (\ref{eq::FUncorr3D})) in the range of radial wavenumber $k^F_{min} < \tilde k_x < k^F_{max}$ and zero forcing covariance matrix outside it.
Figure \ref{fig:NonSteadyStatePoints} shows the evolution of SFH set energy under such external action.
The increasing of the harmonics wavenumber is physically caused by the shear in the unperturbed flow (expression \ref{eq::kx}).
During their evolution, SFHs go through the region of non-zero forcing (denoted by the dashed rectangle in the figure) and here acquire non-zero amplitude (note that harmonics do not interact with each other due to the linearity of the problem and thus their evolution can be calculated independently).
After leaving the forcing region, harmonics evolve as a free SHF taking part in energy exchange processes with the unperturbed flow (see \cite{razdoburdin-zhuravlev-2018} for a brief review).
Finally, all the harmonics under consideration lose their energy by the action of the viscosity.

Now let us take into account that in the flow there exists an infinite number of zero amplitude SFHs with all wavenumbers $-\infty < k_x < \infty$.
Denoting mean energy of all SFHs with wavenumbers in the range from $k$ to $k + dk$ as $E_k$, we found the spectrum of perturbations under stochastic forcing.
We plot the evolution of the spectrum in figure \ref{fig:NonSteadyState} (the initial time moment corresponds to the forcing turning on).
Since the number of zero-amplitude harmonics with $k_x < k^{F}_{min}$ is inexhaustible and forcing is stationary, then steady state spectrum is formed after some relaxation period.
To avoid confusion, wave vector is denoted as $\mathbf{K}$ when we talk about the spectrum of perturbations (instead of $\mathbf{k}$ for wave vector of single SFH).

However, one can calculate the spectrum of perturbations in the steady state much easier.
Let us have a close look at the SFH with the initial value of $\tilde k_x$ equals to the left boundary of the forcing interval: $\tilde k_x(0) = k^F_{min}$ (it corresponds to the far left filled rectangle in figure \ref{fig:NonSteadyState}).
Since after turning on the forcing is stationary, the dynamic of all SFHs having $\tilde k_x(0) < k^F_{min}$ repeats the dynamic of the harmonic having $\tilde k_x(0) = k^F_{min}$, with time shift that equals to $\Delta t = [k^F_{min} - \tilde k_x(0)] / (q k_y)$.
Therefore in moment $t$ the spectrum $E_k(K_x)$ comes to a steady-state for $K_x < K^F_{min} + q K_y t$.
Thus for the calculation of the steady state spectra it is not necessary to integrate equation (\ref{eq::Cdyn}) for numerous SFH as we did above.
It is enough just to integrate equation (\ref{eq::Cdyn}) for SFH with $k_x = k^F_{min}$.
We plot the evolution of such a harmonic in figure \ref{fig:NonSteadyState} by filled rectangles to illustrate that its evolution precisely repeats the steady-state spectrum.

Repeating similar reasoning for the whole of covariance matrix (to avoid confusion we denote spectral density of covariance matrix for steady state as $\mathcal{C}_k$) leads to the following procedure of calculating $\mathcal{C}_k$.
The first step is solving the equation (\ref{eq::Cdyn}) for SFH with $k_x = k^F_{min}$ saving all of the intermediate values of $\mathbf{C}(t)$.
The next step is to find the steady state solution by expression:
\begin{equation}{\label{eq::CSteady}}
\mathcal{C}_k(K_x) = \mathbf{C}\left(\tilde t\right),
\end{equation}
where $\tilde t$ is the moment that corresponds to $\tilde k_x = K_x$ (see expression \ref{eq::kx}).

\begin{figure}
\includegraphics[width=\linewidth]{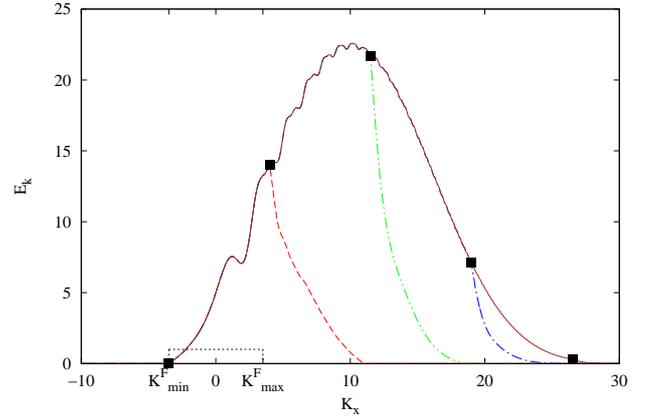}
\caption{
The evolution of perturbations energy spectrum under stochastic forcing for the same parameters as in figure \ref{fig:NonSteadyStatePoints} ($K_y = 1, K_z = 0, R = 1000, R_b = \infty$, $K^F_{min} = -3.5$, $K^F_{max} = 3.5$) is plotted here.
Initially, energy $E_k$ is equal to zero in all range of $K_x$.
Dashed, dot-dot-dashed, dot-dashed and solid lines denote spectrum at moments $t = 5, 10, 15, 20$.
Filled squares denote $E_k$ for SFH which initially has $k_x$ equal to $K^F_{min}$.
The dashed rectangle denotes the spectral interval of forcing.
}
\label{fig:NonSteadyState}
\end{figure}

In contrast to the radial wavenumber $\tilde k_x$, azimuthal and vertical wavenumbers $k_y$, $k_z$ stay permanent during the evolution of SFHs.
Thus in the steady state all the values of $\mathcal{C}_k(K_y, K_z)$ are independent of each other.
At the same time values of $\mathcal{C}_k(K_x)$ are coupled by the equation of SFH dynamics.
For that reason we focus on the characteristics of the steady state integrated over $K_x$ (in the denotation of such integrated quantities we omit the subscript "$k$").
So, integrated over $K_x$ the covariance matrix of the steady-state is 
\begin{equation}
\mathcal{C} = \int\limits_{-\infty}^{\infty} \mathcal{C}_k dK_x,
\end{equation}
spectral energy density integrated over $K_x$ is 
\begin{equation}
E = \mathrm{trace}~\mathcal{C}
\end{equation}
and spectral density of angular momentum flux integrated over $K_x$ is 
\begin{equation}
\Phi = \mathcal{C}_{x, y}.
\end{equation}

In the next section we present an analytical solution for steady state in important subcase of 2D vortical dynamics.
The results of $E$ and $\Phi$ calculations are presented in section \ref{sec::results}.

\end{section}

\begin{section}{Subcase of small scale 2D dynamics}
\label{sec::subcase}

In this section we find an analytical solution of the equation (\ref{eq::Cdyn}) for the case of 2D small scale perturbations.
General assumptions for that subcase consist of a columnar structure of perturbations i.e $k_z = 0$ as well as of incompressible character of the perturbations dynamics.

Equations (\ref{eq::SFH_x} - \ref{eq::SFH_W}) in that case take the following form:
\begin{equation}{\label{eq::SFH2D_x}}
\frac{d \hat u_x}{d t} = 2\hat u_y - {\rm i}\tilde k_x \hat W - R^{-1} k^2 \hat u_x +
    \mathbf{F}_{x, x} \mathbf{\hat\eta}_x + \mathbf{F}_{x, y} \mathbf{\hat\eta}_y,
\end{equation}

\begin{equation}{\label{eq::SFH2D_y}}
\frac{d \hat u_y}{d t} = -(2-q)\hat u_x - {\rm i} k_y \hat W - R^{-1} k^2 \hat u_y + 
    \mathbf{F}_{y, x} \mathbf{\hat\eta}_x + \mathbf{F}_{y, y} \mathbf{\hat\eta}_y
\end{equation}

\begin{equation}{\label{eq::SFH2D_W}}
\tilde k_x \hat u_x + k_y \hat u_y = 0
\end{equation}

Additionally, we should add the limitation of the forcing matrix $\mathbf{F}$, since the expression (\ref{eq::SFH2D_W}) must be valid at every moment in time and the general form of the $\mathbf{F}$ does not conserve incompressible character of perturbations dynamics.
By differentiating the expression (\ref{eq::SFH2D_W}) over $t$
\begin{equation}{\label{eq::divFree}}
\frac{d}{dt}\left(\tilde k_x \hat u_x + k_y \hat u_y\right) = 0
\end{equation}
we get:
\begin{equation}
\tilde k_x \left(\mathbf{F}_{x, x} \mathbf{\hat \eta}_x + \mathbf{F}_{x, y} \mathbf{\hat \eta}_y\right) + 
	   k_y \left(\mathbf{F}_{y, x} \mathbf{\hat \eta}_x + \mathbf{F}_{y, y} \mathbf{\hat \eta}_y\right) = 0.
\end{equation}
Since $\mathbf{\hat \eta}_x$ and $\mathbf{\hat \eta}_y$ are independent noise components, the following pair of equations must be valid independently:
\begin{equation}{\label{eq::F1}}
\tilde k_x \mathbf{F}_{x, x} + k_y \mathbf{F}_{y, x} = 0
\end{equation}
\begin{equation}{\label{eq::F2}}
\tilde k_x \mathbf{F}_{x, y} + k_y \mathbf{F}_{y, y} = 0.
\end{equation}
And, as in expression (\ref{eq::FUncorr3D}), we add a requirement
\begin{equation}{\label{eq::F3}}
\mathrm{trace}~\mathbf{FF}^{\dag} = 1,
\end{equation}
that corresponds to the case of white noise (i.e. spectral density of injected energy does not depend on wavevector).

Conditions (\ref{eq::F1} - \ref{eq::F3}) determine the matrix $\mathbf{FF}^{\dag}$ uniquely:
\begin{equation}{\label{FFdag}}
\mathbf{FF}^{\dag}_{div~free} = 
\begin{pmatrix}
&  k_y^2 / \mathbf{k}^2          & -\tilde k_x k_y / \mathbf{k}^2      \\
& -\tilde k_x k_y / \mathbf{k}^2 &  \tilde k_x^2   / \mathbf{k}^2
\end{pmatrix}.
\end{equation}
Note that since the set of equations (\ref{eq::SFH2D_x} - \ref{eq::SFH2D_W}) has only two independent variables, the matrix $\mathbf{FF}^{\dag}$ has two rows and two columns.
We call such forcing {\it divergence-free forcing} as far as it is obtained from the equation (\ref{eq::divFree}).

For the finding of the covariance matrix in that subcase we use the method of variation of parameters.
First of all, we find a solution for the absence of the external action, i.e. for $\mathbf{FF}^{\dag} = 0$.
The set of equations (\ref{eq::SFH2D_x} - \ref{eq::SFH2D_W}) in the absence of forcing has the analytical solution (see \cite{lominadze-1988} for inviscid case):
\begin{equation}
\hat u_x(t) = u_0 \frac{\sqrt{k_x^2 + k_y^2}}{\mathbf{k}^2} k_y \exp\left[-\gamma(t)\right],
\end{equation}
\begin{equation}
\hat u_y(t) = -u_0 \frac{\sqrt{k_x^2 + k_y^2}}{\mathbf{k}^2} \tilde k_x \exp\left[-\gamma(t)\right]
\end{equation}
with
\begin{equation}
\gamma(t) = \frac{3 k_y^2\left(\tilde k_x - k_x\right) + \left(\tilde k_x^3 - k_x^3\right)}{3 q R k_y}
\end{equation}
and $u_0 = \sqrt{u_x(0)^2 + u_y(0)^2}$ as the initial condition.

The expression for covariance matrix, in that case, has the form:
\begin{equation}{\label{eq::C2D}}
\mathbf{C}_0(t) = u_0^2 \frac{\mathbf{k}_0^2}{\mathbf{k}^4} \exp\left[-2 \gamma(t)\right] 
\begin{pmatrix}
&  k_y^2          & -\tilde k_x k_y \\
& -\tilde k_x k_y &  \tilde k_x^2 
\end{pmatrix}
\end{equation}

The expression (\ref{eq::C2D}) is the solution for equation (\ref{eq::Cdyn}) in the case of 2D incompressible fluid without external forcing.
Now we can find solution for the case of $\mathbf{FF}^{\dag}$ determined by expression (\ref{FFdag}), varying the integration constant $u_0$:
\begin{equation}{\label{eq::C2DF}}
\mathbf{C}(t) = u_0^2(t) \frac{\mathbf{k}_0^2}{\mathbf{k}^4} \exp\left[-2 \gamma(t)\right] 
\begin{pmatrix}
&  k_y^2          & -\tilde k_x k_y \\
& -\tilde k_x k_y &  \tilde k_x^2 
\end{pmatrix}.
\end{equation}

Setting the expression (\ref{eq::C2DF}) to the equation (\ref{eq::Cdyn}) results in the equation for $u_0^2$:
\begin{multline}{\label{eq::C2Df}}
\frac{d u_0^2}{dt} \frac{\mathbf{k}_0^2}{\mathbf{k}^4} \exp\left[-2 \gamma(t)\right] \begin{pmatrix}
&  k_y^2          & -\tilde k_x k_y \\
& -\tilde k_x k_y &  \tilde k_x^2 
\end{pmatrix} = \\
\begin{pmatrix}
&  k_y^2 / \mathbf{k}^2          & -\tilde k_x k_y / \mathbf{k}^2 \\
& -\tilde k_x k_y / \mathbf{k}^2 &  \tilde k_x^2   / \mathbf{k}^2
\end{pmatrix}.
\end{multline}

Simplifying the equation (\ref{eq::C2Df}) we find:
\begin{equation}
\frac{d u_0^2}{dt} \frac{\mathbf{k}_0^2}{\mathbf{k}^2} \exp\left[-2 \gamma(t)\right] = 1.
\end{equation}

This equation has the following solution:
\begin{equation}
u_0^2 = \frac{R}{2 \mathbf{k_0^2}} \exp\left[2 \gamma(t)\right]
\end{equation}

Thus we found the solution for the set (\ref{eq::SFH2D_x} - \ref{eq::SFH2D_W}, \ref{FFdag}):
\begin{equation}
\mathbf{C}(t) = \frac{R}{2\mathbf{k}^4} \begin{pmatrix}
&  k_y^2          & -\tilde k_x k_y \\
& -\tilde k_x k_y &  \tilde k_x^2 
\end{pmatrix}.
\end{equation}

Now we easily find the spectral distribution of covariance matrix in the steady-state with expression (\ref{eq::CSteady}):
\begin{equation}{\label{eq::CAnalyt}}
\mathcal{C}_k(K_x) = \frac{R}{2\mathbf{K}^4} \begin{pmatrix}
&  K_y^2          & -K_x K_y \\
& -K_x K_y &         K_x^2 
\end{pmatrix}.
\end{equation}

The corresponding spectral energy density and spectral angular momentum flux density of induced perturbations are equal to:
\begin{equation}{\label{eq::analytSpectra}}
E_k = \frac{R}{2 \mathbf{K}^2},
\end{equation}
\begin{equation}
\Phi_k = -\frac{K_x K_y R}{2 \mathbf{K}^4}
\end{equation}

The specral density of energy and angular momentum flux of the induced perturbation integrated over $K_x$ are equal to:
\begin{equation}{\label{eq::EAnalyt}}
E = \int\limits_{-\infty}^{\infty} E_k dK_x = \frac{\pi R}{2 K_y},
\end{equation}

\begin{equation}{\label{eq::PhiAnalyt}}
\Phi = \int \limits_{-\infty}^{\infty} \Phi_k dK_x = 0.
\end{equation}

Thus we found that small scale 2D perturbations under stochastic forcing do not produce a flux of angular momentum.
Moreover, the explicit form of the covariance matrix (\ref{eq::CAnalyt}) does not depend on the share rate $q$.
So we found that in the case of small scale 2D dynamics steady-states of perturbations under external stochastic forcing for solid body rotation and non-solid body rotation are equivalent. Wherein in the presence of shear in the flow, the dynamic of single SFH fundamentally differs from the shearless case (SFH in the absence of shear are not liable to transient amplification).
In the next section, we numerically solve the Lyapunov equation for more complicated cases: first, we take compressibility into account and then consider the 3D dynamics of the perturbations.

\end{section}

\begin{section}{Results}
\label{sec::results}

\begin{subsection}{Two-dimensional case}
\label{ssec::results::2D}

In this subsection we analyze the properties of steady-state in the 2D case while taking compressibility into account. First of all, let us make cross-verification of the analytical solution we found in the previous section with the results of numerical integration of the differential Lyapunov equation. Such cross-verification is possible for $K_y \gg 1$, i.e. in the case of small scale perturbations.

Wherein for the correct comparison the covariance matrix of the external forcing should have the same form as was used in the analytical solution (expression (\ref{FFdag})).
Otherwise, external action will induce perturbations with non-zero divergence whose dynamic is not described by the analytical solution we found.
Since the numerical solver is for 3D compressible dynamics, we should pad the matrix $\mathbf{FF}^{\dag}$ with zeros:

\begin{equation}{\label{FFdagCompress}}
\mathbf{FF}^{\dag}_{div~free} = 
\begin{pmatrix}
&  k_y^2 / \mathbf{k}^2          & -\tilde k_x k_y / \mathbf{k}^2  & 0 & 0 \\
& -\tilde k_x k_y / \mathbf{k}^2 &  \tilde k_x^2 / \mathbf{k}^2    & 0 & 0 \\
& 0                              &  0                              & 0 & 0 \\
& 0                              &  0                              & 0 & 0
\end{pmatrix}.
\end{equation}

\begin{figure}
\includegraphics[width=\linewidth]{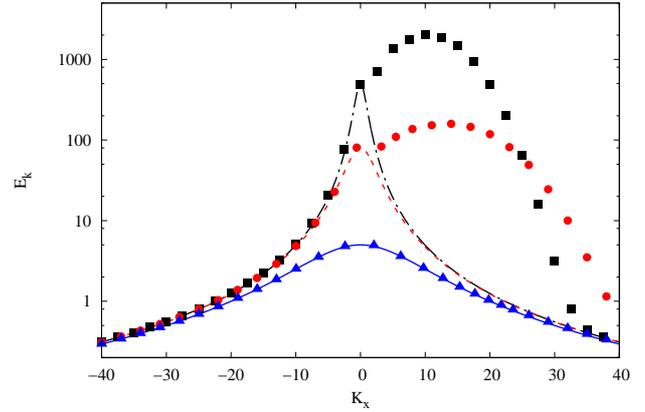}
\caption{
Comparison of analytical and numerical energy density spectra of induced perturabtions for $R = 1000$.
Solid, dashed and dot-dashed lines represent analytical spectra (equation (\ref {eq::analytSpectra})) for $K_y = 1$, $2.5$, $10$ respectively.
Squares, circles and triangles represent numerically found spectra with $R_b = \infty$ for $K_y = 1$, $2.5$, $10$, respectively.
}
\label{fig:E(kx)Comparison}
\end{figure}

\begin{figure}
\includegraphics[width=\linewidth]{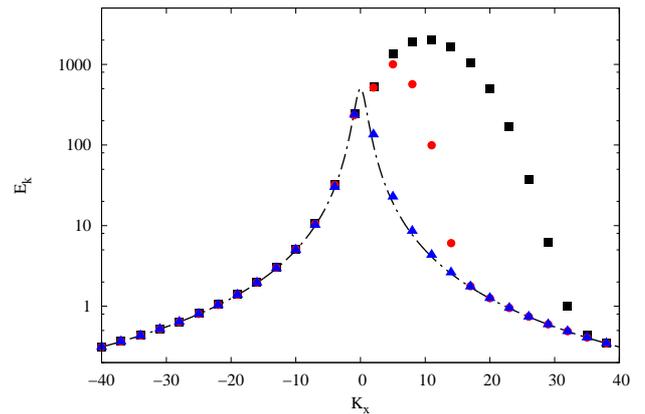}
\caption{
Comparison of analytical and numerical energy density spectra of induced perturbations for $R = 1000$ and $K_y = 1$, $K_z = 0$.
The dot-dashed line represents analytical spectra (equation (\ref {eq::analytSpectra})).
Squares, circles and triangles represent numericaly found spectra for $R_b = \infty$, $100$, $1$, respectively.
}
\label{fig:E(kx)ComparisonRb}
\end{figure}
Moreover, we should note that we found the analytical solution for unlimited spectral diapason of the forcing. The numerical solver cannot reproduce that feature exactly since it should have finite limits of integration. However, if the limits are quite large, their subsequent increasing would not change the result. 

Insofar as the illustrative comparison of covariance matrices is impossible, we limit ourselves to the comparison of energy density spectra.
In figure \ref{fig:E(kx)Comparison} we plot $E_k(K_x)$ both for analytical (expression \ref{eq::analytSpectra}) and numerical solutions for different values of azimuthal wavenumber.
One can see that the difference between analytical and numerical spectra becomes negligible for $K_y \gg 1$.
Thus we conclude that the two solutions are cross-verified for the case of small-scale perturbations.
Moreover, one can note that, when two solutions differ from each over, the difference occurs in the region of positive radial wavevectors $K_x > 0$.
This feature is the key to understanding the physical reason for the difference.
As we discussed in section \ref{sec::spectra}, the spectrum of induced perturbations is connected to the evolution of single SFH by expression (\ref{eq::CSteady}).
Thus we conclude that the difference between analytical and numerical solutions occurs at the swing interval of SFH (region of $\tilde k_x \sim 0$).
At this interval, transiently amplified vortex acquires the properties of sound wave, and it becomes impossible to separate one from another.
As a result, the vortices emit sound waves (see \cite{chagelishvili-1997}, \cite{bodo-2005}, \cite{heinemann-papaloizou-2009a}).

To check this hypothesis, we vary the bulk-viscosity by decreasing the corresponding Reynolds number $R_b$.
Bulk viscosity acts only on sound waves but not on vortices.
Thus, if our interpretation is correct, the difference between the two solutions should decrease with the decreasing of $R_b$.
In figure \ref{fig:E(kx)ComparisonRb} we plot the analytical spectrum for $K_y = 1$ and numerical spectra for the same $K_y$ with different values of $R_b$. 
One can find the difference disappear for $R_b = 1$.
That perfectly fits the proposed interpretation.

The covariance matrix of the forcing that is determined by the expression (\ref{FFdagCompress}) does not cover all the possibilities in the case of a 2D compressible fluid. 
Matrix $\mathbf{FF}^{\dag}$ for the case of uncorrelated forcing (2D analog of expression \ref{eq::FUncorr3D})
\begin{equation}{\label{eq::FUncorr2D}}
\mathbf{FF}^{\dag}_{uncorr} = 
\begin{pmatrix}
& 1/2 & 0   & 0 & 0 \\
& 0   & 1/2 & 0 & 0 \\
& 0   & 0   & 0 & 0 \\
& 0   & 0   & 0 & 0
\end{pmatrix}
\end{equation}
is equal to half-sum of (\ref{FFdagCompress}) and another matrix:
\begin{multline}{\label{FFdagRotFree}}
\mathbf{FF}^{\dag}_{rot~free} = 
\begin{pmatrix}
&  \tilde k_x^2 / \mathbf{k}^2   & \tilde k_x k_y / \mathbf{k}^2 & 0 & 0 \\
& \tilde k_x k_y / \mathbf{k}^2  & k_y^2 / \mathbf{k}^2   & 0 & 0 \\
& 0                              & 0                      & 0 & 0 \\
& 0                              & 0                      & 0 & 0  
\end{pmatrix}. 
\end{multline}
We call the forcing determined by such a matrix {\it rotor-free forcing} since one can derive it from saving the irrotational character of perturbation velocities. 

Due to the linearity of the equation (\ref{eq::Cdyn}), one can represent covariance matrix of the steady-state $\mathcal{C}_k$ for the forcing determined by the matrix (\ref{eq::FUncorr2D}) as a half-sum of solutions found for divergence-free and rotor-free forcing matrices.
In figure \ref{fig:E(kx)RotFree} we plot the comparison of induced perturbations spectra for the divergence-free forcing with the rotor-free one.
The comparison of energy $E$ and angular momentum flux $\Phi$ integrated over $K_x$ for the cases of divergence-free and rotor-free forcing are plotted in figure \ref{fig:EF(ky)}.
Despite of equal forcing power,
\begin{equation}{\label{eq::rotFree}}
\mathrm{trace}~\mathbf{FF}^{\dag}_{div free} = \mathrm{trace}~\mathbf{FF}^{\dag}_{rot free} = 1,
\end{equation}
the total induced perturbations energy for divergence-free forcing significantly exceeds that of the rotor-free case for all values of azimuthal wavevector $K_y$.
Wherein, $\Phi$ for the divergence-free forcing quickly decreases during the increasing of $K_y$, $\Phi$ for the rotor-free case, however, stays constant. 

\begin{figure}
\includegraphics[width=\linewidth]{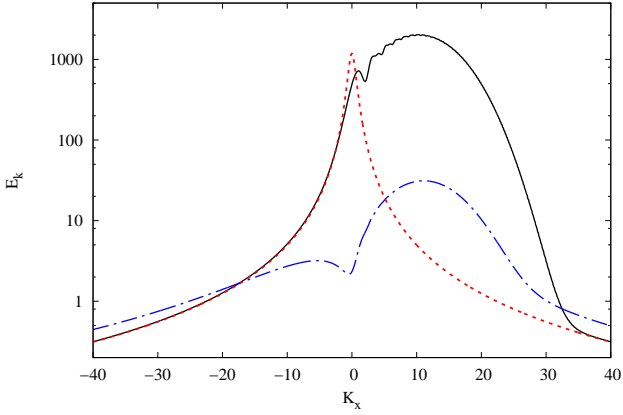}
\caption{
The comparison of induced perturbations energy density spectra for divergence-free forcing (solid line), rotor-free forcing (dot-dashed line) both for $K_y = 1$, $K_z = 0$, $R = 1000$, $R_b = \infty$.
The analytical solution (expression \ref{eq::analytSpectra}) for the same parameters is denoted by the dashed line.
}
\label{fig:E(kx)RotFree}
\end{figure}

\begin{figure}
\includegraphics[width=\linewidth]{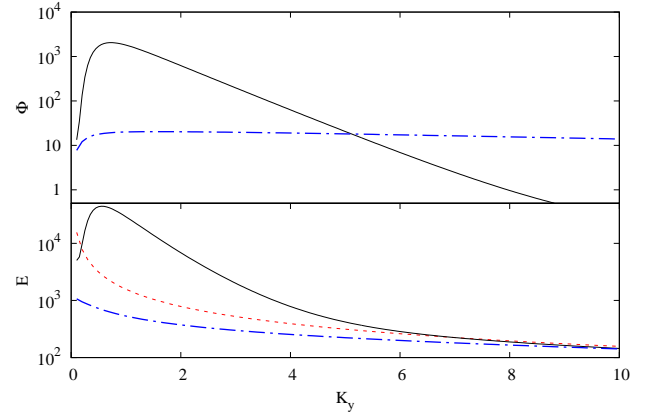}
\caption{
On the top and bottom panels spectral density of angular momentum flux and energy integrated over $K_x$ are plotted.
Both are in the in steady state for $K_z = 0$, $R = 1000$, $R_b = \infty$.
The dashed line represents analytical solutions for small-scale dynamics (expression (\ref{eq::EAnalyt}).
Solid and dot-dashed lines represent numerical solutions with forcing determined by expression (\ref{FFdagCompress}) and (\ref{FFdagRotFree}) respectively.
}
\label{fig:EF(ky)}
\end{figure}

Thus for the 2D compressible dynamics with external forcing, there are two mechanisms of angular momentum transfer to the flow's  periphery.
The first one dominates for perturbations with large azimuthal scale.
The action of that mechanism is the following:
\begin{itemize}
\item Divergence-free component of the forcing generates shearing vortices.
\item These vortices do not provide angular momentum transfer by themselves, but their amplitude significantly increases by the swing amplification.
\item At the swing interval (that corresponds to the maximum of these amplitudes) vortices emit sound waves that provide positive angular momentum flux.
\end{itemize}
The second mechanism dominates for small-scale perturbations, for which the emission of sound waves by vortices is suppressed.
This mechanism operates due to the generation of shearing sound waves by the rotor-free component of the forcing.
The waves transfer angular momentum as such, but the magnitude of the flux is much smaller than in the case of large-scale perturbations.

The last question we want to touch upon for the 2D case is the role of spectral localization of the forcing.
We have previously considered that external action operated at the whole diapason of radial scales.
Now we set the covariance matrix of the forcing $\mathbf{FF}^{\dag}$ non-zero only in certain diapason of radial wavenumbers: $K_{min}^F < K_x < K_{max}^F$.
This case is better physically motivated than the previous one since now the injected power
\begin{equation}
\dot E_{in} = 
\int\limits_{-\infty}^{\infty} \dot e_{in} dK_x = 
\int\limits_{K_{min}^F}^{K_{max}^F} \mathrm{trace}~\mathbf{FF}^{\dag} dK_x
\end{equation}
is finite.

We denote the size of spectral diapason of the forcing as $\Delta K^F = K_{max}^F - K_{min}^F$.
The center of that diapason is denoted as $K_x^F = (K_{min}^F + K_{max}^F) / 2$.
Similarly, we denote azimuthal wavenumber of the forcing as $K_y^F$.
As far as we consider (see the explanation before equation \ref{eq::FUncorr3D}) that the trace of the covariance matrix of the forcing is equal to the unit, injected power numerically equals to $\Delta K_x$.

\begin{figure}
\includegraphics[width=\linewidth]{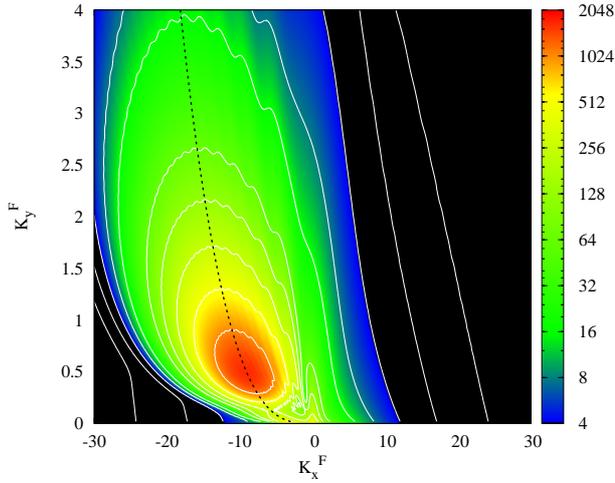}
\caption{
Colormap of $E / \dot E_{in}$ for different spectral diapasons of forcing.
Dotted line shows analytically predicted values of optimal $K_x$ as function of $K_y$ (equation \ref{eq::kx_max}).
Solid lines denote iso-levels equal to the power of $2$.
Covariance matrix of the forcing is determined by the expression (\ref{eq::FUncorr2D}).
$R = 1000$, $R_b = \infty$, $K_z = 0$, $\Delta K = 0.02$.
}
\label{fig::Map:Ekxky:R=1e3:kz=0}
\end{figure}

\begin{figure}
\includegraphics[width=\linewidth]{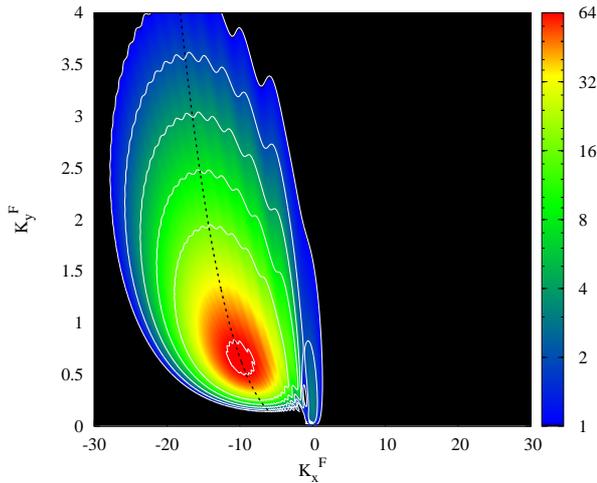}
\caption{
Colormap of $\Phi / \dot E_{in}$ for the same parameters as in figure \ref{fig::Map:Ekxky:R=1e3:kz=0}.
The dotted line shows analytically predicted values of optimal $K_x$ as function of $K_y$ (equation \ref{eq::kx_max}).
Solid lines denote iso-levels equal to the power of $2$.
}
\label{fig::Map:Fkxky:R=1e3:kz=0}
\end{figure}

As far as injected power is finite now, we can use the relation of the energy of perturbations or angular momentum flux integrated over $K_x$ provided by them in steady-state to the injected power ($E / \dot E_{in}$ and $\Phi / \dot E_{in}$ respectively).
The relations are convenient since they demonstrate the effectiveness of forcing.
We plot $E / \dot E_{in}$ and $\Phi / \dot E_{in}$ for varying $K_y^F$ and $K_x^F$ in figures \ref{fig::Map:Ekxky:R=1e3:kz=0} and \ref{fig::Map:Fkxky:R=1e3:kz=0}.
In those figures, one can easily see spectral diapasons in which the external action is most effective.

Previously we found that swing amplification of the vortices plays an essential role in the shaping of steady-state.
Thus we mark an optimal for swing relation between $K_x^F$ and $K_y^F$
(see estimations for the maximum transient growth in \cite[section 5]{afshordi-2005}):
\begin{equation}{\label{eq::kx_max}}
K_{x, opt}^F = -\left(q R K_y^F\right)^{1/3}.
\end{equation}
by the dotted line in figures \ref{fig::Map:Ekxky:R=1e3:kz=0} and \ref{fig::Map:Fkxky:R=1e3:kz=0}.
One can make sure that the maximuma of both $E / \dot E_{in}$ and $\Phi / \dot E_{in}$ in the figures are close to that line.

\end{subsection}

\begin{subsection}{Three-dimensional case}
\label{ssec::results::3D}

In this subsection we investigate the steady-state of perturbations under external stochastic forcing in the 3D case.
The main interest is to find certian 3D mechanisms that can provide angular momentum transfer and to check how the efficiency of the 2D one will degrade in presence of vertical non-homogeneousness.

Our approach for both these goals is calculating $E / \dot E_{in}$ and $\Phi / \dot E_{in}$ as function of $K_y^F$ and $K_z^F$ (see figures \ref{fig::Map:Ekykz:R=1e3:kx=opt} and \ref{fig::Map:Fkykz:R=1e3:kx=opt}).
At the same time, we associate $K_x^F$ and $K_y^F$ by expression (\ref{eq::kx_max}).
With such approach the mechanism of angular momentum transfer that differs from the 2D one manifests itself in the form of local maximum with $K_z^F \ne 0$.

\begin{figure}
\includegraphics[width=\linewidth]{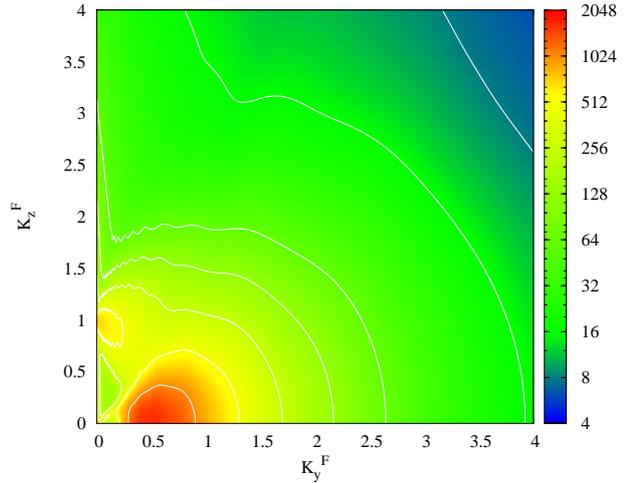}
\caption{
Colormap of $E / \dot E_{in}$ for different spectral diapasons of forcing. 
Solid lines denote iso-levels equal to the power of $2$.
Covariance matrix of the forcing is determined by the expression (\ref{eq::FUncorr3D}).
$R = 1000$, $R_b = \infty$, $\Delta K = 0.02$, $K_x^F = K_{x, opt}^F$ (equation \ref{eq::kx_max}).
}
\label{fig::Map:Ekykz:R=1e3:kx=opt}
\end{figure}

\begin{figure}
\includegraphics[width=\linewidth]{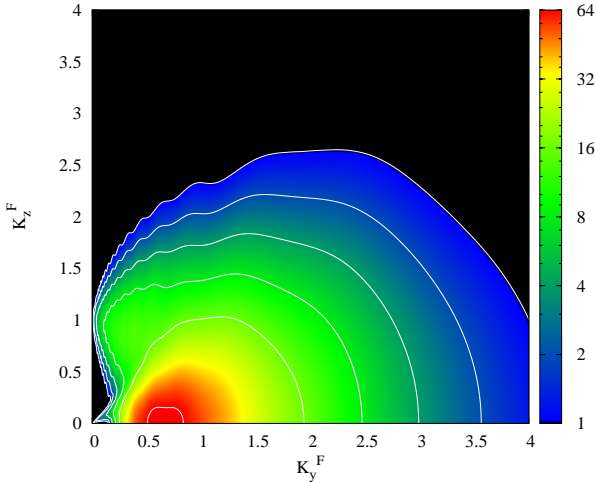}
\caption{
Colormap of $\Phi / \dot E_{in}$ for the same parameters as in figure \ref{fig::Map:Ekykz:R=1e3:kx=opt}.
Solid lines denote iso-levels equal to the power of $2$.
}
\label{fig::Map:Fkykz:R=1e3:kx=opt}
\end{figure}

One can find that the effectiveness of the 2D mechanism decreases with the increasing of $K_z^F$.
On the other hand, induced perturbation still provides angular momentum flux up to $K_z^F \sim 1$.
So the 2D mechanism we discussed above stays significant even for vertical scales comparable to disk thickness.

Moreover by a close look at figure \ref{fig::Map:Ekykz:R=1e3:kx=opt} one can find a local maximum of $E / \dot E_{in}$ for $K_z^F\sim 1$, $K_y^F \ll 1$.
For a more detailed investigation of the process resulting in that local maximum, we plot energy and angular momentum spectra of induced perturbations for the corresponding azimuthal and vertical wavenumbers (see figure \ref{fig::EF::kz=1}).
Dot-dashed curves in the figure correspond to the absence of the bulk-viscosity in the flow.
As in the 2D case, maxima of the spectra have been shifted at the region of positive radial wavenumbers.
Moreover the comparison of the spectra for $R_b = \infty$ and $R_b = 0.1$ demonstrate that the difference is most prominent in the region of $K_x > 0$.
Comparing these features with the ones for the 2D case we assume that for SFHs with $k_z \sim 1$ these exists the process of density waves emission that is a 3D analogue of the 2D emission (see \cite{chagelishvili-1997}, \cite{bodo-2005}, \cite{heinemann-papaloizou-2009a}).
A more detailed investigation of the process is beyond the goals of the current paper.
Thus, we leave this subject for future research.

\begin{figure}
\includegraphics[width=\linewidth]{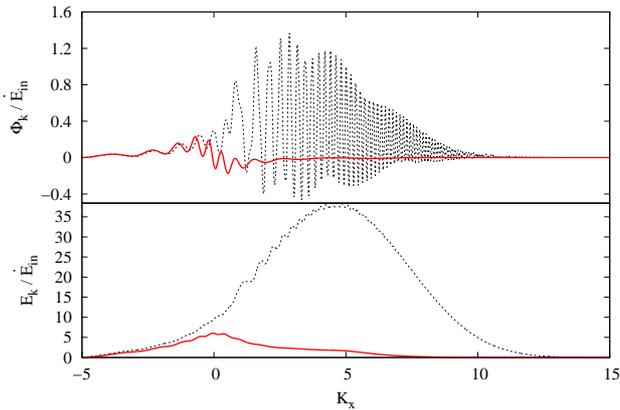}
\caption{
Spectra of induced perturbations energy (bottom panel) and angular momentum flux (top panel).
Solid and dashed lines correspond to $R_b = 0.1$ and $R_b = \infty$, respectively. $K_z^F = 1$, $K_y^F = 0.1$, $K_{min}^F = -5$, $K_{max}^F = 5$, $R = 1000$.
}
\label{fig::EF::kz=1}
\end{figure}

\end{subsection}

\begin{subsection}{Dependence on the Reynolds number}
\label{ssec::results::Reynolds}

In the previous subsections we focused on the physical interpretation of the mechanisms giving rise to angular momentum transfer by linear perturbations.
To make the comparison clear, we fixed Reynolds number to be $R = 10^3$ in all the antecedent calculations.
However, in real accretion flows, Reynolds numbers can exceed $10^{10}$.
Thus, the dependence of the effectiveness of the mechanisms on Reynolds number is critically essential.
For that reason in the current subsection we calculate the dependence of $E / \dot E_{in}$ and $\Phi / \dot E_{in}$ on $R$ (see figure \ref{fig::EFOptimal}).

\begin{figure}
\includegraphics[width=\linewidth]{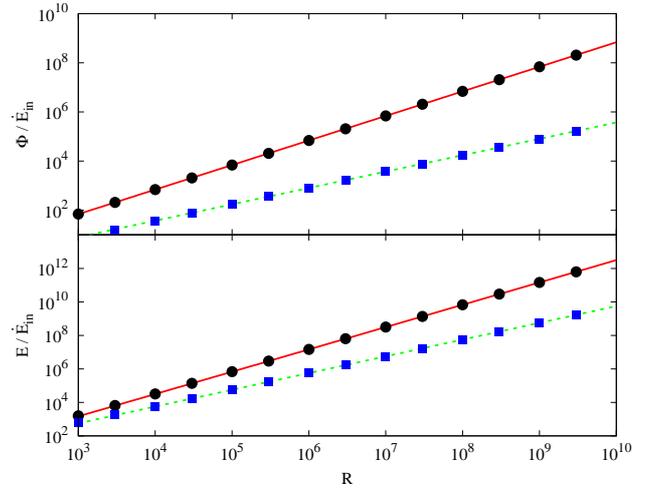}
\caption{
Integrated over $K_x$ energy of induced perturbations under external stochastic forcing (bottom panel) and angular momentum flux provided by those perturbations (top panel) plotted vs Reynolds number.
The cicrcles correspond to the 2D case with $K^F_z = 0$, $K_y^F = 0.6$, $R_b = \infty$, $K^F_x = K^F_{x, opt}$ (eq. \ref{eq::kx_max}) and forcing covariance matrix determined by the expression (\ref{eq::FUncorr2D}).
The solid lines power functions that provide an optimal fit of the circles.
The rectangles correspond to the 3D case with $K^F_z = 1$, $K^F_y = 0.1$, $R_b = \infty$, $K^F_x = K^F_{x, opt}$ (eq. \ref{eq::kx_max}) and the forcing covariance matrix determined by the expression (\ref{eq::FUncorr3D}).
The dashed lines are power functions that provide the optimal fit of the rectangles.
}
\label{fig::EFOptimal}
\end{figure}

Since we found that for both cases the dependencies are power-law-like, we also make the fitting of the exponents (least squares algorithm was used
\footnote{We use Gnuplot (\cite{gnuplot}) for least squares fitting as well as for plotting of all the figures).}
).
\begin{itemize}
\item For the 2D case ($K_z^F = 0$, $K_y^F = 0.6$), the dependencies are
\begin{equation}{\label{eq::approx2D}}
E / \dot E_{in} \sim R^{4/3},~~\Phi / \dot E_{in} \sim R.
\end{equation}

\item For the 3D case ($K_z^F = 1$, $K_y^F = 0.1$), the dependencies are:
\begin{equation}
E / \dot E_{in} \sim R,~~\Phi / \dot E_{in} \sim R^{2/3}.
\end{equation}
\end{itemize}

One can find that both the angular momentum transfer produced by the induced perturbations and their energy fastly increase with Reynolds number.
Thus, since typical Reynolds numbers in accretion disks can exceed $10^{10}$, external action with even small amplitude can provide significant transfer of angular momentum to the periphery of the flow.
\end{subsection}

\end{section}

\begin{section}{Summary}{\label{sec::summary}}

In the current paper, we investigated the steady-state of linear perturbations that arise in local compressible Keplerian flow under external stochastic forcing.
We do not touch upon the source of the external action, as far as for a variety of accretion flows such sources can be completely different.
Wherein, if certain natural limitations are imposed on the properties of the forcing, the steady-state of the perturbations can be described without taking the source of forcing in consideration.
Here is the list of these limitations:
\begin{enumerate}
\item The external force is spatially $\delta$-correlated, temporally Gaussian stochastic process with zero ensembles mean.

\item Stochastic addition to the continuity equation is zero.

\item The ensemble averaged covariance matrix of the forcing is time-independent.
\end{enumerate}

The main goal of the paper consisted in investigating the angular momentum transfer by the induced perturbations in the steady state.
We found several mechanisms that can give rise to the angular momentum transfer to the periphery of the flow.
The most powerful of them is based on the swing amplification of induced vortices with the followed emission of density waves.
In spite of the fact that this scenario is based on the 2D processes, it remains effective even for vertical scales comparable to disk thickness.

We also found that if the injected power is fixed, the amount of transferred angular momentum is a linear function of the Reynold number.
That makes the proposed scenario important for accretion and protoplanetary disks which are characterized by enormous Reynold numbers. 

Thus an external action with even small amplitude can provide significant angular momentum transfer to the periphery of the flow. 
Wherein, the mechanism of the flux providing is independent of the boundary conditions and global flow structure.
For that reason, we suggest that the scenario under investigation become considered as an additional source of angular momentum flux in accreting flows.

Moreover, stochastic forcing can be looked upon as the reason for the effective viscosity fluctuations that is required in the model of \cite{lybarskii-1997} for describing the flicker noise oscillations.
\end{section}

\section*{Acknowledgements}
The equipment for the reported study was granted by the M. V. Lomonosov Moscow State University Programme of Development.

\appendix
\begin{section}{Derivation of equation for dynamic of covariance matrix}
\label{appendix::C::derivation}
The main idea of equation derivation is the same as used in \cite{farrell-ioannou-1996a}, \cite{farrell-ioannou-1999}, \cite{ioannou-kakouris-2001} for autonomous operators. However, the time-dependence of the operator leads to change in some details of the derivation procedure.

First of all, let us write the solution for the homogeneous form of equation (\ref{eq::dynamic}) (see section 2 of \cite{farrell-ioannou-1996b}):

\begin{equation}{\label{eq::homogenSolution}}
\mathbf{q}(t) = \lim_{\delta t \to 0} \prod \limits_{j = 1}^{n} \exp \left[\mathbf{A}(t_j) \delta t\right] \mathbf{q_0},
\end{equation}
here $t = t_0 + n \delta t$, $t_0 + (j - 1) \delta t < t_j < t_0 + j \delta t$, $\mathbf{q_0} = \mathbf{q}(t_0)$.
The production of the infinite small propagators is marked as 
\begin{equation}
\mathbf{B}(t, t_0) = \lim_{\delta t \to 0} \prod \limits_{j = 1}^{n} \exp \left[\mathbf{A}(t_j) \delta t\right].
\end{equation}
The inverse propagator is equal to
\begin{equation}
\mathbf{B}^{-1}(t, t_0) = \lim_{\delta t \to 0} \prod \limits_{j = n}^{1} \exp\left[-\mathbf{A}(t_j) \delta t\right]
\end{equation}
Further, the following properties of the propagator will be required:
\begin{equation}
\frac{d \mathbf{B}(t, t_0)}{dt} = \mathbf{A}(t) \mathbf{B}(t, t_0)
\end{equation}
and for $t > s > t_0$
\begin{equation}
\mathbf{B}(t, t_0)\mathbf{B}^{-1}(s, t_0) = \mathbf{B}(t, s).
\end{equation}

Substituting solution (\ref{eq::homogenSolution}) into inhomogeneous equation (\ref{eq::dynamic}) we get:
\begin{equation}
\frac{d \mathbf{q}_0}{d t} = \mathbf{B}^{-1}(t, t_0) \mathbf{F}(t)\mathbf{\eta}(t).
\end{equation}

Thus
\begin{equation}
\mathbf{q}_0(t) = \int\limits_{t_0}^t \mathbf{B}^{-1}(s, t_0) \mathbf{F}(s) \mathbf{\eta}(s) ds
\end{equation}
and
\begin{equation}
\mathbf{q}(t) = \int\limits_{t_0}^t \mathbf{B}(t, s) \mathbf{F}(s) \mathbf{\eta}(s) ds
\end{equation}

Since we have an explicit solution, the derivation of expression for the covariance matrix (\ref{eq::Cdef}) does not provide any difficulties:
\begin{multline}
\mathbf{C}_{ij}(t) = \mathbf{C}_{ij}(0) + \Bigg\{ \Bigg<\int\limits_0^tds\int\limits_0^tds^{\prime} \mathbf{B}(t, s)\mathbf{F}(s)\mathbf{\eta}(s) \times \\ 
\mathbf{\eta}^{\dag}(s^{\prime})\mathbf{F}^{\dag}(s^{\prime}) \mathbf{B}^{\dag}(t, s)\Bigg>\Bigg\}_{ij}.
\end{multline}
Here $\mathbf{C}_{ij}$ are components of the matrix $\mathbf{C}$.

After taking the equation (\ref{forcing_covariance}) into account, we get the final expression for $\mathbf{C}$:
\begin{multline}{\label{C_expr}}
\mathbf{C}_{ij}(t) = \mathbf{C}_{ij}(0) + \Bigg\{\int\limits_0^t \mathbf{B}(t, s) \mathbf{F}(s)\mathbf{F}^{\dag}(s) \mathbf{B}^{\dag}(t, s) ds\Bigg\}_{ij}.
\end{multline}

By the differentiation of expression (\ref{C_expr}) by $t$ we get the equation (\ref{eq::Cdyn}):
\begin{equation}
\frac{d\mathbf{C}}{dt} = \mathbf{F}\mathbf{F}^{\dag} + \mathbf{AC} + \mathbf{CA}^{\dag}
\end{equation}
\end{section}

\begin{section}{Adjoint dynamical operator}
\label{appendix::Adag::derivation}
Here we present the derivation for the adjoint operator $\mathbf{A}^{\dag}$.
The technique of derivation is the same as has been used in \cite[section 6.3.4]{razdoburdin-zhuravlev-2018}.

First of all we write a scalar product of the two state vectors, $\mathbf{f}$ and $\mathbf{g}$.
\begin{equation}{\label{eq::scalar_prod}}
\left(\mathbf{f}, \mathbf{g}\right) = f_x g_x^* + f_y g_y^* + f_z g_z^* + f_w g_w^*,
\end{equation}
where subscripts $x, y, z, w$ correspond to the components of the perturbation.
It is easy to note that the norm generated by that scalar product represents the acoustic energy of the perturbation.

We decompose operator $\mathbf{A}$ into three components
\begin{equation}{\label{eq::A}}
\mathbf{A} = \mathbf{D} + \mathbf{S} + \mathbf{B},
\end{equation}
where $\mathbf{D}$ corresponds to the action of non-viscous terms, $\mathbf{S}$ -- to the action of kinematic viscosity, $\mathbf{B}$ -- to the bulk viscosity.

An explicit view of $\mathbf{D}$, $\mathbf{S}$ and $\mathbf{B}$ matrix representations can be easily obtained from the equation (\ref{eq::dynamic}) and the set (\ref{eq::SFH_x} -- \ref{eq::SFH_W}):
\begin{equation}{\label{eq::D}}
\mathbf{D} = 
\begin{pmatrix}
& 0,                   & 2,            & 0,             & -{\rm i} \tilde k_x \\
& -(2 - q),            & 0,            & 0,             & -{\rm i} k_y \\
& 0,                   & 0,            & 0,             & -{\rm i} k_z \\
& -{\rm i} \tilde k_x, & -{\rm i} k_y, & -{\rm i} k_z,  & 0 \\
\end{pmatrix}
\end{equation}

\begin{equation}{\label{eq::S}}
\mathbf{S} = -R^{-1} 
\begin{pmatrix}
& \mathbf{k}^2, & 0,            & 0,            & 0 \\
& 0,            & \mathbf{k}^2, & 0,            & 0 \\
& 0,            & 0,            & \mathbf{k}^2, & 0 \\
& 0,            & 0,            & 0,            & 0 \\
\end{pmatrix}
\end{equation}

\begin{equation}{\label{eq::B}}
\mathbf{B} = -\left(R^{-1} / 3 + R_b^{-1}\right)
\begin{pmatrix}
& \tilde k_x \tilde k_x, & \tilde k_x k_y, & \tilde k_x k_z, & 0 \\
& \tilde k_x k_y,        & k_y k_y,        & k_y k_z,        & 0 \\
& \tilde k_x k_z,        & k_z k_y,        & k_z k_z,        & 0 \\
& 0,                     & 0,              & 0,              & 0 \\
\end{pmatrix}
\end{equation}
Now we can get the explicit form of adjoint equations directly from the definition:
\begin{equation}
\left(\mathbf{f}, \mathbf{A g}\right) = \left(\mathbf{A}^{\dag} \mathbf{f}, \mathbf{g}\right).
\end{equation}
with help of equation (\ref{eq::scalar_prod}).

\begin{equation}{\label{eq::Adag}}
\mathbf{A}^{\dag} = \mathbf{D}^{\dag} + \mathbf{S} + \mathbf{B},
\end{equation}
with 
\begin{equation}{\label{eq::Ddag}}
\mathbf{D}^{\dag} = 
\begin{pmatrix}
& 0,                  & -(2 - q),    & 0,            & {\rm i} \tilde k_x \\
& 2,                  & 0,           & 0,            & {\rm i} k_y \\
& 0,                  & 0,           & 0,            & {\rm i} k_z \\
& {\rm i} \tilde k_x, & {\rm i} k_y, & {\rm i} k_z,  & 0 \\
\end{pmatrix}.
\end{equation}
Here we take into account that $\mathbf{S}$ and $\mathbf{B}$ are self-adjoint.
\end{section}

\begin{section}{Testing of equation solver without stochastic forcing}
\label{appendix::solver::test}
\begin{figure}
\includegraphics[width=\linewidth]{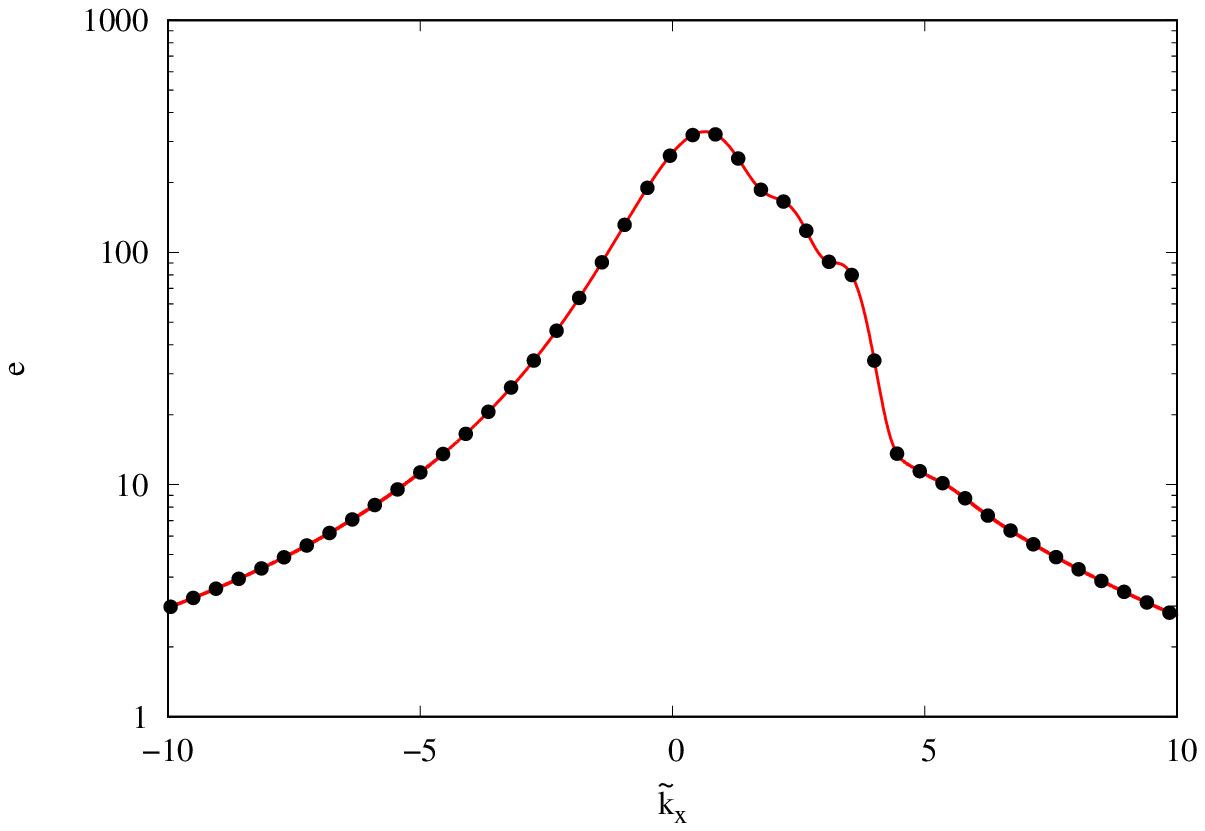}
\caption{
The energy of single SFH plotted vs radial wavenumber.
The points represent the numerical solution of the set (\ref{eq::SFH_x} -- \ref{eq::SFH_W}) with {\bf the} initial conditions (\ref{eq::IC_x} -- \ref{eq::IC_W}).
The solid line represents the numerical solution of matrix equation (\ref{eq::Cdyn}) with the initial condition (\ref{eq::IC_C}).
Both solutions are obtained for the same set of parameters: $R = 10000$, $R_b = 4$, $k_y = 1$, $k_z = 0$, $k_x = -20$.
}
\label{fig::e(kx)Free}
\end{figure}

Here we examine our numerical solver for matrix equation (\ref{eq::Cdyn}) comparing it with the solution for the set of equations for the dynamics of single SFH (see equations (\ref{eq::SFH_x} -- \ref{eq::SFH_W})).
If the external action is absent (case of $\mathbf{F} = 0$), two solutions should reproduce each other for the same initial conditions.

For our test, we choose the initial conditions in the form of leading spirals with the initial norm equal to the unit:

\begin{equation}{\label{eq::IC_x}}
\hat u_x(0) = \frac{k_y}{\sqrt{k_x^2 + k_y^2}}
\end{equation}

\begin{equation}
\hat u_y(0) = -\frac{k_x}{\sqrt{k_x^2 + k_y^2}}
\end{equation}

\begin{equation}{\label{eq::IC_W}}
\hat W(0) = 0.
\end{equation}

{\bf This} corresponds to the following initial covariance matrix:
\begin{equation}{\label{eq::IC_C}}
\mathbf{C}(0) = 
\begin{pmatrix}
&  \frac{k_y^2}{k_x^2 + k_y^2}   & -\frac{k_x k_y}{k_x^2 + k_y^2} & 0 & 0 \\
& -\frac{k_x k_y}{k_x^2 + k_y^2} &  \frac{k_x^2}{k_x^2 + k_y^2}   & 0 & 0 \\
&   0                            &  0                             & 0 & 0 \\
&   0                            &  0                             & 0 & 0
\end{pmatrix}
\end{equation}

For an illustrative comparison, we use the energy of the SFH (see equation \ref{eq::e}).
By varying the problem parameters ($R$, $R_b$, $k_y$ and $k_x$) and comparing the energy of the SHF we made sure that the two  solutions are equivalent (see figure \ref{fig::e(kx)Free} for illustration).

Note once again that everywhere outside this paragraph the initial condition for covariance matrix equals to zero.
\end{section}

\end{document}